\documentclass[conference, 10pt]{IEEEtran}
\ifCLASSINFOpdf
\else
\fi
\usepackage{graphicx}
\usepackage{color}
\usepackage{placeins}
\usepackage{float}
\usepackage{hyperref}
\usepackage{tabularx,colortbl}

\begin{document}
\title{Parallel-Plate Waveguides Formed by Arbitrary Impedance Sheets}


\author{\IEEEauthorblockN{X.~Ma\IEEEauthorrefmark{1}\IEEEauthorrefmark{2}}
\IEEEauthorblockA{\IEEEauthorrefmark{1}Department of Electronics and Information\\
Northwestern Polytechnical University\\
Xi'an, Shaan Xi, China\\
maxin1105@nwpu.edu.cn\\}
\and
\IEEEauthorblockN{M.~S.~Mirmoosa\IEEEauthorrefmark{2}  S.~A.~Tretyakov\IEEEauthorrefmark{2}}
\IEEEauthorblockA{\IEEEauthorrefmark{2}Department of Electronics and Nanoengineering\\
Aalto University\\
Espoo, Finland\\}}

\maketitle

\begin{abstract}
%\boldmath
In this work, we introduce and study parallel-plate-waveguide structures that are formed by two penetrable metasurfaces having arbitrary sheet impedances. We investigate the guided modes which can propagate in such structures and derive the corresponding dispersion relations. Different scenarios including series- and parallel-resonant impedance sheets are considered. The obtained theoretical formulas are applied to predict the dispersion properties for different separations between the two symmetric or asymmetric metasurfaces. 
\end{abstract}


% \fi
\IEEEpeerreviewmaketitle

\section{Introduction}
Waveguides have many applications in microwave and optical technologies. Following the development of artificial impedance surfaces~\cite{FSS1,tretyakov}, novel waveguide structures that exploit impedance surfaces have been investigated for guiding and controlling waves~\cite{sievenpiper1}--\cite{F}. Such waveguides, which were initially considered for the purpose of radiating electromagnetic energy, are supporting  surface waves.
Surface waves along a single sheet~\cite{sievenpiper1} or along a sheet placed over a ground~\cite{maci2} have been studied  and exploited for controlling the propagation path of guided utilizing stop-bands or for leaky-wave radiation~\cite{grbic2}. However, most of the studies, from the early days till today, have been focused on guided waves in one sheet and little is known about what happens if we bring another thin sheet close to the first one. Can we modify and control the waveguide properties if we use two parallel sheets?

In this presentation, we will discuss our results regarding these two sheets which are parallel to each other and separated by a distance. We study guided modes supported by such structures assuming that the sheets have arbitrary isotropic impedances. We also reveal the electric currents density at the sheets for different polarizations and distances between those two sheets. 

\section{Properties of the structure under study}
\begin{figure}[t!]
	\begin{center}
		\noindent
		\includegraphics[width=3in]{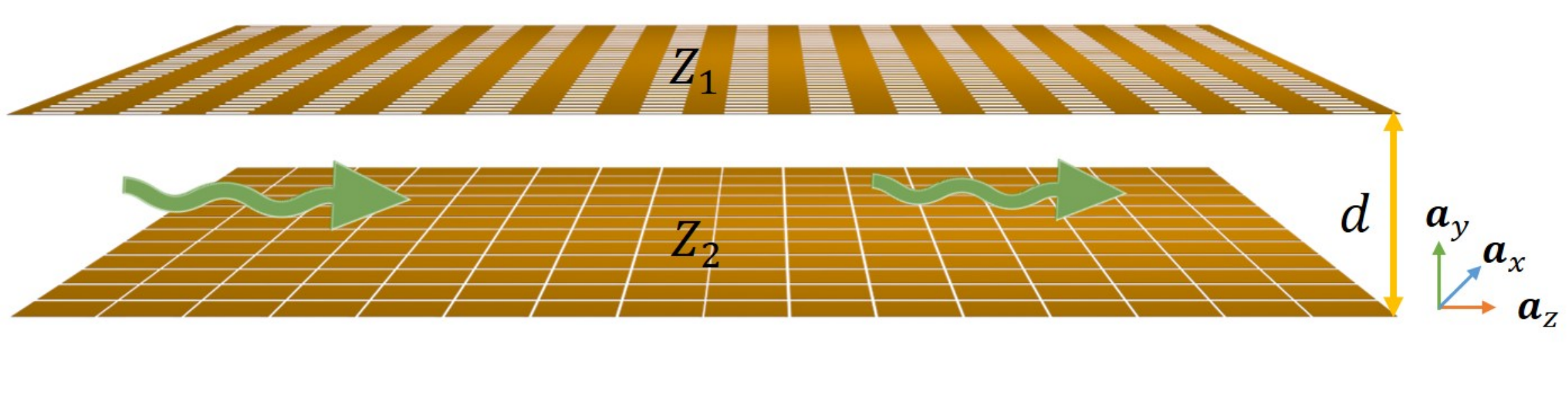}
		\caption{The structure under study: Two parallel impedance sheets. The wave is propagating along the $\mathbf{a}_z$-direction.}\label{fig:st}
	\end{center}
\end{figure}

\begin{figure}[t!]
	\begin{center}
		\noindent
		\includegraphics[width=2.6in]{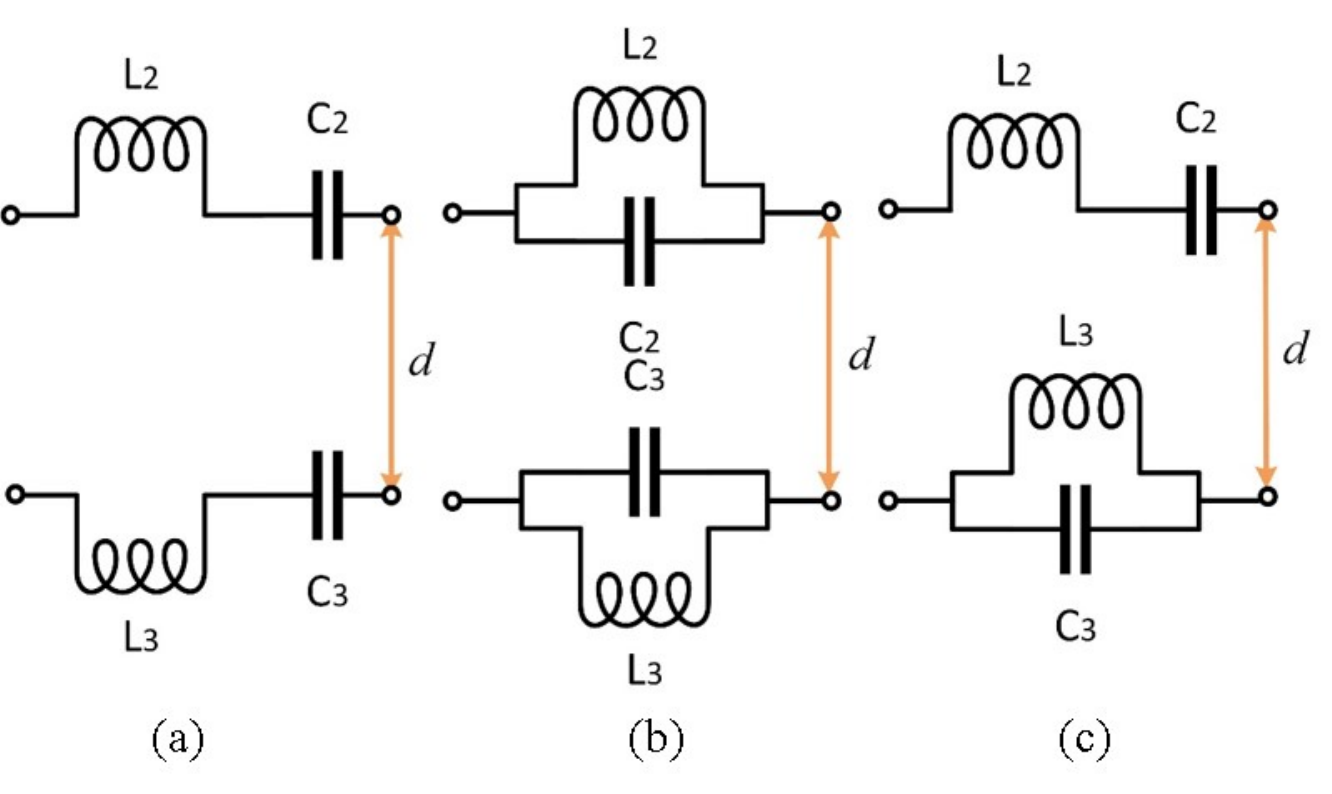}
		\caption{The three scenarios for resonant-impedance waveguides in which (a) the impedances of both sheets are series connections, (b) the impedances of both two sheets are parallel connections and (c) the impedance of one sheet is a series connection and another one is a parallel connection.}\label{fig:circuit}
	\end{center}
\end{figure}
The proposed and studied parallel-plate waveguide is shown in Fig.~\ref{fig:st}. It consists of two parallel sheets which are separated by the distance $d$ and have only electric response. One sheet is supposed to be positioned at $y=d/2$ and consequently the other one is placed at $y=-d/2$. The space between the two sheets is filled by air (vacuum). We assume that the electromagnetic wave is propagating along the $\mathbf{z_0}$-axis. The impedances of the sheets are denoted as $Z_1$ and $Z_2$, respectively. Figure~\ref{fig:circuit} illustrates the realization of the impedance of each sheet as a series or parallel connection of effective sheet inductance ($L$) and capacitance ($C$). This means that the equivalent impedance of each sheet has a resonance at $\omega_{\rm{res}}$=$1/\sqrt{LC}$. Since we have two sheets, there can be three different scenarios as shown in Fig.~\ref{fig:circuit}. Regarding the first and second scenarios, the impedances of both two sheets are realized by a  series or parallel connection, respectively. For the third scenario, one sheet is realized by a series connection and the other one by a parallel connection. For each scenario, two different cases can be considered: Asymmetric and symmetric. In the  asymmetric case, the two sheets possess different impedances ($Z_1\neq Z_2$) while in the symmetric case the sheets are exactly the same ($Z_1=Z_2$).

\begin{figure}[t!]
	\begin{center}
		\noindent
		\includegraphics[width=3.5in]{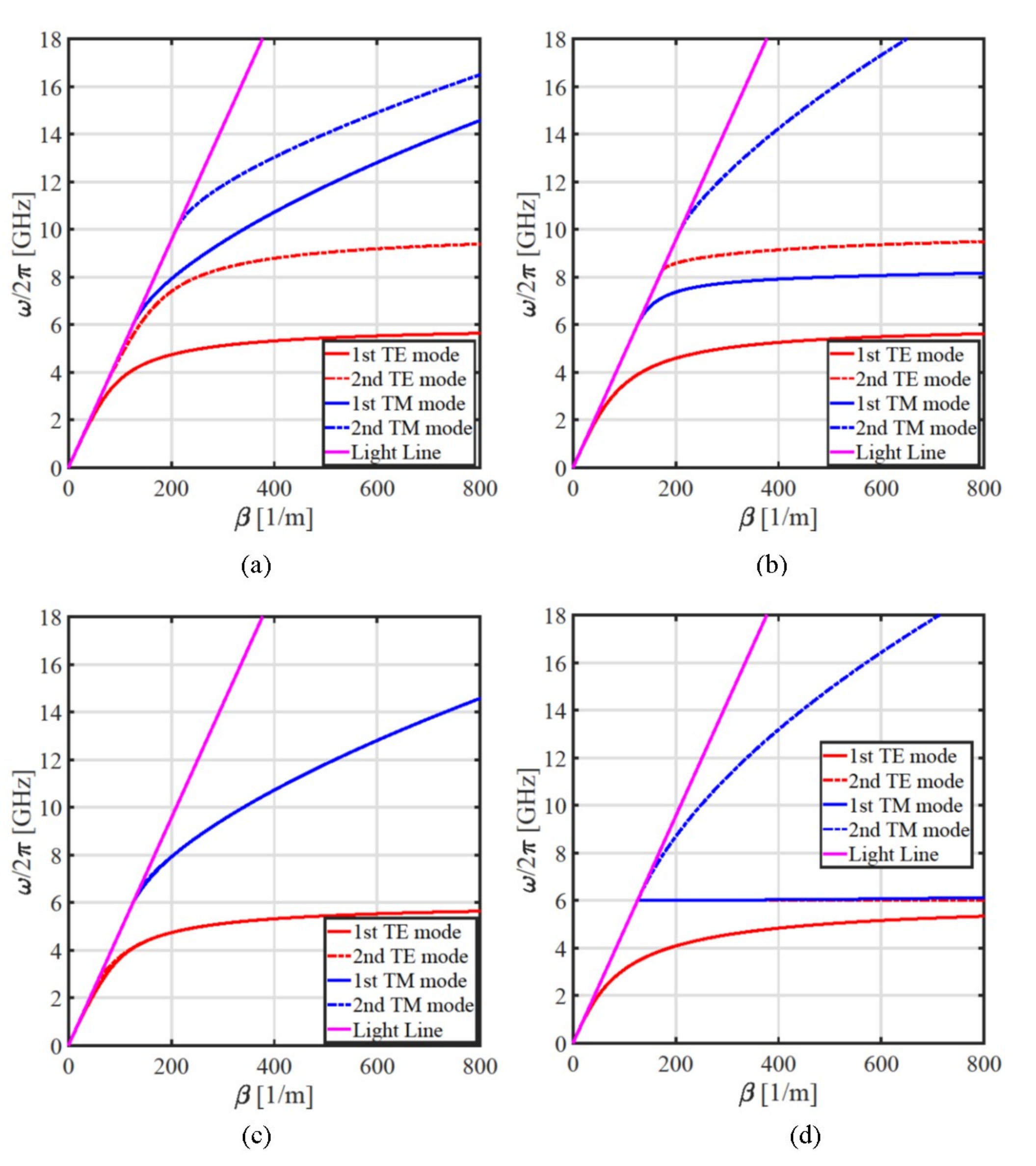}
		\caption{Dispersion curves of the first scenario in which  the two sheets are  (a) asymmetric when $d=\lambda_{\rm{6GHz}}$, (b) asymmetric when $d=\lambda_{\rm{6GHz}}/5000$,  (c) symmetric when $d=\lambda_{\rm{6GHz}}$ and (d) symmetric when $d=\lambda_{\rm{6GHz}}/5000$.}\label{fig:different series}
	\end{center}
\end{figure}

\section{Results and Discussion}

There is a possibility to decompose the fields into two polarizations: The transverse magnetic polarization (TM modes) and the transverse electric polarization (TE modes). The polarizations of propagating modes depend on the impedance of the sheets. If both sheets are inductive, two TM modes can propagate within the structure. In the case of two capacitive sheets, the two modes have the TE polarization. Both polarizations can simultaneously contribute to transferring the energy if one sheet is inductive and the other one is capacitive.

Here, let us focus on only one scenario (the series connection) and discuss the dispersion diagrams corresponding to the asymmetric and symmetric cases. Results for  the other scenarios and cases will be presented in the talk. As an example, we suppose that the resonance frequency of the sheets are  $f_{\rm{res}}^1=6$~GHz and $f_{\rm{res}}^2=10$~GHz, respectively, for the asymmetric case, and $f_{\rm{res}}^1=f_{\rm{res}}^2=6$~GHz for the symmetric case.  The analytical results for the dispersion curves of the first scenario are presented in Fig.~\ref{fig:different series}. In the asymmetric case, when the distance is equal to $\lambda_{\rm{6GHz}}$, two TE modes are supported below $6$~GHz since both sheets are capacitive. In the range from $6$ to $10$~GHz, one sheet is capacitive and the other sheet is inductive, which gives rise to propagation of waves of both polarizations. Above $10$~GHz, two TM modes can propagate because both sheets are inductive. It is worth to mention that the two TM modes have cut-off frequencies because of the resonance, while one of the TE modes suffers from a cut-off frequency which depends on the separation. As the distance between the sheets decreases, the cut-off frequency of the second TE mode increases. When $d$ is small but still not exactly zero, a new resonance frequency $f_{\rm{mix}}$ emerges, and the dispersion curves are  determined by $f_{\rm{res}}^1$, $f_{\rm{res}}^2$, and $f_{\rm{mix}}$. When $d=0$, $f_{\rm{mix}}$ is the same as  the cut-off frequency of the second TE mode. Interestingly, a very narrow stop band arises at $f_{\rm{mix}}$ for surface waves when $d$ is exactly equal to zero. 

Clearly, the symmetric sheet arrangement corresponds to a special case of the asymmetric structure with $f_{\rm{res}}^1=f_{\rm{res}}^2=f_{\rm{mix}}$. In the symmetric case, one can see that two TE modes are supported below $6$~GHz because the two sheets are capacitive, and  two TM modes can propagate above $6$~GHz since two sheets are inductive, when the distance between the two sheets is comparable with $\lambda_{\rm{6GHz}}$. As we bring the two sheets close to each other, the first TM mode and the second TE mode are compressed to the resonance frequency. Such dual-polarized resonant mode supporting waves with a wide range of propagation constants can have applications in super-resolution imaging devices.
% Only one TE mode and one TM mode can propagate within such structure below and above the resonance frequency, respectively.

\section{Conclusion}
In this paper, we have introduced and studied a novel wave-guiding structure,formed by two metasurfaces which can have arbitrary reactive sheet impedances. We have theoretically studied  guided modes of symmetric and asymmetric structures for different distances between the sheets. Generally, the topology of dispersion diagrams relates to the resonance frequencies. In the limit of zero distance between the two sheets, a new resonance frequency emerges and four modes are classified by three resonance frequencies within the asymmetric structure, while for the symmetric case only two modes can propagate and only one resonance frequency exist. In the presentation, we will also give physical insight into the behavior of modes based on the current density and distribution of fields and discuss the other scenario and cases as well.

\section*{ACKNOWLEDGEMENT}
This work was supported in part by the Aeronautical Science Foundation of China (No. 20161853018) and the Basic Science Foundation (No. 2017205B020).

\end{document}